\pdfoutput=1

\documentclass[10pt]{article} 
\usepackage{ragged2e}
\usepackage[utf8]{inputenc} 
\usepackage[pdftex]{graphicx}
\usepackage{amssymb} 

\usepackage{geometry} 
\usepackage{graphicx} 
\usepackage{amsmath}
\usepackage{hyperref} 
\usepackage{mathtools}
\usepackage{upgreek}
\usepackage[parfill]{parskip} 

\usepackage{booktabs} 
\usepackage{array} 
\usepackage{paralist} 
\usepackage{verbatim} 
\usepackage{subfig} 

\usepackage{fancyhdr} 
\usepackage{sectsty}
\allsectionsfont{\sffamily\mdseries\upshape} 

\usepackage[nottoc,notlof,notlot]{tocbibind} 
\usepackage[titles,subfigure]{tocloft} 

\title{PCA-based Data Reduction and Signal Separation Techniques for James-Webb Space Telescope Data Processing}
\author{Hatipoğlu, Güray}

\begin{document}
\maketitle

\centering{\section*{Abstract}}
\justifying
\noindent Principal Component Analysis (PCA)-based techniques can separate data into different uncorrelated components and facilitate the statistical analysis as a pre-processing step. Independent Component Analysis (ICA) can separate statistically independent signal sources through a non-parametric and iterative algorithm. Non-negative matrix factorization is another PCA-similar approach to categorizing dimensions in physically-interpretable groups. Singular spectrum analysis (SSA) is a time-series-related PCA-like algorithm. After an introduction and a literature review on processing JWST data from the Near-Infrared Camera (NIRCam) and Mid-Infrared Instrument (MIRI), potential parts to intervene in the James Webb Space Telescope imaging data reduction pipeline will be discussed.

\setlength{\columnsep}{3em}
\setlength{\columnseprule}{1pt}

\centering\section{Methodological Background}
\justifying

\centering\subsection{PCA}
\justifying

Readers can look at the tutorial of Shlens (2014) and the book of Everitt and Hothorn (2011) to comprehend the theory behind PCA and its advantages/shortcomings. Below is a brief introduction to PCA: \\
\textit{Covariance matrix construction}\\
We have a dataset with N samples and M variables. Firstly, the mean of each variable will be subtracted from its corresponding sample value for every sample. The purpose behind this covariance matrix and using covariance is to see how much the initial variables vary together. The formula for this covariance is as follows:

\[
    cov_{xy} = \sigma_{xi}\sigma_{yi}= 
    \sum_{i=1}^{n} (x_{i} - \mu_{x})*(y_{i} - \mu_{y})
\]
\\
The variables that \textit{vary together} will have higher covariance values, resulting from the multiplication of two values diverged from their mean together, compared to the cases where one variable varies but the other is near its mean.
\\
\noindent After this step, constructing the correlation matrix permits checking the multicollinearity among variables. A matrix decomposition method (e.g. eigenvalue decomposition) can be applied to this matrix. The entire dataset will be rotated from the original variables to a new space with \textit{principal components} ordered from the maximum variance in data to the lowest one. Principal components are linear combinations of variables with different weights. For instance, the first PC is:
\[
     \alpha_{1}^{'}\textbf{x} = \alpha_{11}\textit{x}_{1} +  \alpha_{12}\textit{x}_{2} + ... +  \alpha_{1p}\textit{x}_{p} + = 
    \sum_{i=1}^{m} (\alpha_{1i}\textit{x}_{i})
\]
\\
The second PC is needed to be the one with the highest variance included in it after PC1 was extracted from the dataset, and it should be orthogonal to PC1 and all other PCs. Further details are in Joliffe (2002).
Eigenvalue decomposition is one way to generate PCs (eigenvectors) and a set of eigenvalues that imply the total variance loaded on each PC. A square matrix (like the one we will have after generating the covariance matrix above) will be eigen-decomposed as below:

\centering
\[
A = Q* 
\lambda
*Q^{-1}
\]
\\
\justifying
\noindent where A above can be our covariance matrix generated above. $\lambda$ at the middle of the right-hand's side of the equation is a diagonal matrix with its diagonals containing eigenvalues, and the Q is an eigenvector matrix (see Shlens (2014) for a tutorial on PCA). This is the classical PCA. One issue with it is that its components are difficult to interpret. A \textit{varimax} rotation, for instance, might ease the interpretability by reducing the number of moderate loadings and increasing 1,0,-1 (Kaiser, 1958), yet this is generally associated with a trade-off in the explanatory power of the components.  In addition to the several methods outlined below, a \textit{sparsity} tweak can potentially ameliorate this problem. One example of the construction of a sparse PCA is present in Zou, Hastie, and Tibshirani (2006).

\centering\subsection{Independent Component Analysis-ICA}
\justifying

Independent Component Analysis (ICA) goes one step further than PCA (or singular value decomposition) and aims to generate \textit{statistically independent} components, in addition to the zero correlation. Shlens (2014b) provided a tutorial on independent component analysis with theory, examples, and sample code to run an efficient ICA algorithm. ICA has two steps, removing second-order correlations via SVD, then removing the remaining (assumed) correlations numerically with statistical techniques. SVD is as follows:

\centering
\[
A = U* 
\Sigma
*V^{T}
\]
\\
\justifying
\noindent U is the left singular vector matrix, V is the right singular vector matrix, and $\Sigma$ is a diagonal matrix with singular values in decreasing order from top left to right bottom. Since U and V matrices are orthogonal, they assist in proceeding with the decomposition of matrix A in the following way:

\[
A = U* 
\Sigma
*V^{T}, A^{T} = V* 
\Sigma^{T}
*U^{T}
\]

\[
A^{T}A = V* 
\Sigma^{T}
*U^{T} U* 
\Sigma
*V^{T} = V* 
\Sigma^{T}\Sigma
*V^{T}
\]

\[
AA^{T} = U* 
\Sigma
*V^{T} V* 
\Sigma^{T}
*U^{T} =  U* 
\Sigma^{T}\Sigma
*U^{T}
\]

\noindent At this point, we solve both $A^{T}A$ and $AA^{T}$ similar to the way in eigendecomposition since they are structurally similar. Moreover, the multiplication of the transpose of U and V with themselves results in identity matrix I. According to the method in Shlens (2014b), half of the decomposition (U and $\Sigma$) is solved through a PCA-like approach for second-order correlation removal with the covariance of the data as input. Then, the remaining V matrix is numerically found to ensure the singular components have the minimum dependence on each other. Statistical independence:

\centering
\[
I(y) = 
\int {P(y) \log_{2}{\frac{P(y)}{\prod_{i} P(y_{i})}      } dy}
\]
\justifying
This is the definition for multi-information, where the non-negative $I(y)$ value is at its lowest (0 value) if and only if all $y_{i}$ are independent of each other. After the first PCA step for decorrelation and multiplication with singular values (\textit{normalization}):

 \centering
\[
\hat{s} = Vx_{w} 
\]
\justifying

\noindent where $\hat{s}$ is the source matrix, $x_{w}$ is the matrix after previous transformations, and V is the remaining rotation matrix. The critical issue here is as follows: Normally, we will return to A when U, $\Sigma$, and V are multiplied in this order. However, constructing V in such a way that maximizes the independence between the resulting components will give us different components in the inverse of the columns of the resulting operation ($W^{-1}$):

\centering
\[
W = V*D^{\frac{-1}{2}}*E^{T} 
\]
\justifying

\noindent where W is the \textit{unmixing} matrix. Up to now, there is no noise assumption, i.e., the components are the actual components of the data, such as the voice of one person at the cocktail party and the background music. In this way, ICA tries to generate components statistically independent from each other. One of its extensions, \textit{dependent component analysis} (Li, Li, and Wang, 2010), can find statistically independent clusters of components, where each cluster itself has dependent components in it.

\centering\subsection{Non-negative Matrix Factorization-NMF}
\justifying
Non-negative matrix factorization (NMF) is one way to decompose a matrix into non-negative components, which facilitates interpretability.
\[
M = UV^{T} 
\]
\justifying M is the original matrix with n rows and p columns (m x n), U is (m x r), V is (n x r), with the target as:
\[
\arg \min_{\{U,V\}\geq{0}}{||M-UV^{T}||_{F}}
\]
\justifying There are numerous computationally efficient algorithms to retrieve non-negative entries from the matrix M with different assumptions. One example is separable non-negative matrix factorization (SNPA). Its additional separability assumption can be defined iteratively until \textit{r} rank (Nadisic et al., 2006) with a \textit{J} index as below:

\[
{H}^{*}(:,j) = \arg \min_{h\epsilon \Delta}{f(M(:,j)-M(:,J)h)}
\]

\[
{R(:,j) = M(:,j) - M(:,J){H}^{*}(:,j)}
\]

\centering\subsection{SSA}
\justifying

Singular spectrum analysis (SSA) is suitable for time-series data. It does not make \textit{a priori} assumptions about the data, and it is a non-parametric method. It has many different branches, and the main form (one-dimensional) has the following steps (Hassani, 2010) \\
1-] Trajectory matrix:\\
With an \textbf{L} window length, one-dimensional time-series data becomes multi-dimensional \textit{Hankel matrix} as: \\

\[
{X} = 
\begin{bmatrix}
y_{1}&y_{2}&y_{3}& \cdots &y_{K}\\
y_{2}&y_{3}&y_{4}& \cdots&y_{K+1}\\
y_{3}&y_{4}&y_{5}& \cdots&y_{K+2}\\
\vdots &\vdots&\vdots &\ddots& \vdots\\
y_{L}&y_{L+1}&y_{L+2}&\cdots &y_{T}
\end{bmatrix}
\]

\justifying
The skew-diagonal elements are equal in this Hankel matrix. \\
2-] Singular Value Decomposition of \textit{XX}$^{T}$: \\
This will be in the form of $\textit{XX}^{T}$ = $P\Lambda P^{T}$ \\
3-] Eigen-vector Selection: \\
This is the part where we separate the one-dimensional series into different components. There are many ways to deal with this step, and one of them is grouping correlated elements into one cluster, and separating the most uncorrelated ones in different clusters (Golyandina and Korobeynikov, 2013), looking at the correlation between the reconstructed elements, (w-correlation matrix). Furthermore, if there is an \textit{a priori} knowledge about the frequency of a specific signal or noise, choosing \textit{L} window as a multiple of that frequency will assist in separating that component. More details regarding the recommendations over selecting \textit{L} window length are present in Golyandina, Nekrutkin, and Zhigljavsky (2001). \\
4-] Reconstruction of the One-dimensional time series: \\
This is via $PP^{T}$\textbf{X} after selecting components of X above in step 3.

\centering\section{Literature Review}
\justifying

\centering\subsection{Near-Infrared Camera-NIRCam}
\justifying
Commissioning and First On-Sky Results of the JWST/Near Infrared Camera-NIRCam, with the characteristics of the NIRCam itself can be found in Girard et al. (2022).\\
Schlawin et al. (2022) examined the performance of the near-infrared camera (NIRCam) short-wavelength time-series photometry with JWST data on the HAT-P-14 system. They reported that the additional errors from \textbf{mirror tilt events} are within 50 \% of the theoretical photon noise. The frequency of these events declines gradually but has not disappeared completely yet. Among several other methods, they also utilized PCA to sense tilt events with NIRCam grism and fine guidance sensor (FGS), yet the results were nearly on the noise level. Their best method was \textit{noise-weighted differential dot product}:\\

\centering
\[
{D}_{i} = ((\vec{A}_{i}-\vec{R})/\vec{V}) \cdot (\vec{M}-\vec{R}) 
\]
\justifying
$\vec{A}_{i}$ is single FGS cal data image, $\vec{R}$ is the reference image after tilt, $\vec{M}$ is the reference image before tilt and $\vec{V}$ is the variance image. This method required additional information on when the tilt event occurred from NIRCam data, as well. These tilts might be present in all cycle 1 observations.\\
Ahrer et al. (2022) first studied WASP-39 with Next-Generation Transit Survey (NGTS) and Terrestrial Exoplanet Searching Satellite (TESS) to ensure that WASP-39b is a suitable target to monitor with JWST. After this step, 8.2 hours of observation with NIRCam took place with Grism R + F322W2 filter in long wavelength (LW, 2.420-4.025 $\mu$m) channel, and WLP8 weak lens and F210M filter in short wavelength (SW, 2.0-2.2 $\mu$m). Both modes used the SUBGRISM256 subarray mode and SHALLOW4 readout pattern. In the end, 366 integrations were available for the transit observation. They utilized the standard data reduction pathway in \textbf{jwst} Python package with minor deviations. In \textit{short-wavelength photometry}, Eureka! and tshirt pipelines were utilized. After aperture photometry in Eureka!, the target aperture radius was 65 pixels, with background annulus between 70 and 90 pixels relative to the center. \textbf{tshirt} had row-by-row, odd/even-by-amplifier (ROEBA) subtraction to reduce 1/f noise. The photometric extraction had a source radius of 79 pixels, with 79-100 pixels as the background annulus. In \textit{long-wave spectroscopy}, Eureka!, atmospHeric trANsmission SpectrOscopy anaLysis cOde (HANSOLO), tshirt, and chromatic-fitting pipelines were used. For reference, commissioning program 1076 with a third-degree polynomial wavelength solution from Pfund and Bracket Hydrogen Series of planetary nebula IRAS 05248-7007. Eureka! Stage 1 and 2, with jump detection setting at 6$\sigma$, were the same as the standard \textbf{jwst} pipeline. After several filtering steps, they fitted the spectral trace to a Gaussian profile. Background subtraction had a double-iteration 7$\sigma$ threshold for outlier rejection. HANSOLO began with the jwst stage 1 calibrated outputs. Cosmic ray effects were removed according to the LACOSMIC algorithm and the Moffat function identified the spectral trace and fitted it to each column. The sky was removed after a fitted linear trend for each column and its subtraction, excluding the region of 20 pixels on either side of the trace center. Different images' spectra were aligned with cross-correlation. In tshirt Stage 3, optimal spectral extraction had covariance between pixels as weights, with further spatial exclusion details in the paper's methodology section. They cautioned about the large error deviations in the edges, so recommends the utilization of 2.420-4.025 $\mu$m wavelength region. They select their slightly modified Eureka! as the best, and they conducted atmospheric radiative-convective equilibrium models of ATMO, PHOENIX, and PICASO 3.0. All their forward models produced the solar to the super-solar-metallicity atmosphere (1-100 x solar), sub-stellar C/O ratio ($\leq$0.36), and substantial cloud contribution. They have further constraints regarding ${H}_{2}{O}, {CH}_{4}$, and ${CO}_{2}$ abundances.\\
Carter et al. (2022), along with MIRI, which is summarized in the MIRI subsection below, utilized NIRCam of JWST on HIP 65426 b super Jupiter exoplanet for High Contrast Imaging (HCI) purposes. They employed angular differential imaging (ADI), and also reference differential imaging (RDI) with HIP 68245. These reference observations had nine separate dither positions to construct a library of PSF capturing a different misalignment between the star and coronagraphic mask with each science exposure. spaceKLIP python package generated PSF subtracted images, contrast curves, and companion photometry and astrometry. NIRCam coronagraphy has the MASK335R round coronagraphic mask with two roll angles consecutively for F250M, F300M, F410M, F356W, and F444W filters. The authors cautioned against the erroneous "jump" detections in the NIRCam MULTIACCUM detector readout. For both NIRCam and MIRI data, PSF was subtracted with three different PCA-based methods as ADI, RDI, and ADI+RDI.

\centering\subsection{Mid-Infrarent Instrument-MIRI}
\justifying
Bouwman et al. (2022) made a mid-infrared time-series observation of L 168-9 b, a transiting exoplanet, with JWST Mid-Infrared Instrument (MIRI) data. The time-series observation (TSO) lasted for 4.14 hours with 9371 integrations (with the PID 1033 code Observation 5 in Mikulski Archive for Space Telescopes (MAST) archive for data). The mode was slitless low-resolution spectroscopy (LRS), F1000W filter with a successful target acquisition step. The pointing stability was excellent according to the result of the fine guidance sensor (FGS) and High-Gain Antenna (HGA) engineering telemetry data. Their best calibration route with Detector1 portion of \textit{jwst} pipeline for raw data processing had the following steps: dq init, saturation, reset, linearity, last frame, dark current, jump step with 5.0 modified threshold, and ramp fitting; with their details elaborated on their paper. In the second stage, spectral extraction and time-series analysis, they employed two methods that largely have their bases on default \textbf{jwst} Spec2Pipeline. JWST Calibration Reference Data System (CRDS) pipeline mapping version 859 was used for calibration and reference files. \textbf{assign wcs} and \textbf{flat field} assigned RA and DEC to every pixel in the spectral images. For background correction, a time-dependent approach was followed and this median background was subtracted separately for each integration. The median background was cross-dispersion for 10 tı 17 and 57 to 72 detector columns. They used CASCADe-filtering from Calibration of transit Spectroscopy using the Casual Data (CASCADe) package for cosmic hit and bad pixel search. Then, CASCADe-jitter located the spectral trace, with 3x3 Scharr-Operator calculated (Scharr, 2000) Jacobian and Hessian matrices in Canny edge filter (Canny, 1986) implementation. This step made second-order Taylor expansion of the Hessian matrix in the direction of the maximum value eigenvector for the spectral trace-related pixels according to the Canny edge filter. The spectra in 1 dimension came from \textbf{extract1d} pipeline step with custom parameters came from the trace fit generated above in the previous step. Later, the CASCADe transit spectroscopy package calibrated the time-series data and extracted the transmission spectrum of L 168-9 b through a half-sibling regression methodology. They rebinned the spectra with a spectral resolution of 50 at 7.5 $\mu$m, and the first 30 minutes of time-series data, corresponding to 1019 integrations were skipped as they have strong response drifts. After setting eccentricity and inclination from L 168-9 b discovery (Astudillo-Defru et al., 2020), observed mid-transit time in data, and relative semi-major axis, limb darkening correction was non-linear by Claret (2000) with ExoTETHyS package (Morello et al. 2020). The lightcurve calculation was with the Batman package (Kreidberg, 2015). Errors in their transit depth fit and systematics were estimated with a bootstrap sampling having 600 samples, with more details in Carone et al. (2021). Besides the CASCADe pathway, they also used 0.5 version \textbf{Eureka!} pipeline (Bell et al., 2022) in 5 different stages. Stage 1 outputs were the same as CASCADe, while stage 2 has the 11.16.5 version of the \textbf{crds} package and 1.6.0 of \textbf{jwst} pipeline. The skipped "photom" step to prevent its degradation on time-series data. Stage 3 noise removal window had 140-393 y-rows and 13-64 x-rows. A source aperture half-width of 4 and a background exclusion region half-width of 10 after considering 8-16 were chosen. Stage 4 binned the spectra in 48 channels with 0.149 $\mu$m widths in 4.86-1.86 $\mu$m and a white-light channel. The HGA-caused anomaly was also removed in this step with a box-car filter of 500 integrations in width with a maximum of ten iterations using an iterative sigma clipping. Stage 5 employed \textbf{dynesty} to fit the observations according to the \textbf{batman} transit model, with further details on the paper on the white-light curve and limb-darkening-related processes. In the end, the results of both CASCADe and Eureka! paths were in agreement within themselves and with Patel \& Espinoza (2022). \\
Besides NIRSpec, Miles et al. (2022) studied brown dwarf VHS 1256 b with MIRI on JWST, too. MIRI had all 4 IFU channels from 4.98 to 28.1 $\mu$m with short, medium, and long grating for around 2 hours. The read-out mode was FASTR1. Background subtraction in the JWST pipeline was not used, instead, a reference aperture was placed off of the target in our calibrated science cubes. Similar to the NIRSpec, collapse and 2D Gaussian fit was conducted prior to the aperture photometry. ResidualFringeStep detected and removed fringes that remained in the portions of the MIRI spectrum. The entire Channel 4 from 17.71 to 20.94 $\mu$m collapsed to a single photometric point. They reported that the coincided wavelength range between NIRSpec and MIRI has consistent results. In the end, they reported disequilibrium chemistry with turbulent vertical mixing, small silicate particles capable of forming clouds, with the presence of $H_{2}O$, $CH_{4}$, $CO$, and $CO_{2}$.\\
Carter et al. (2022) also utilized MIRI along with NIRCam coronagraphy for HIP 65426 b with a four-quadrant mask coronograph, with each quadrant tied to a different filter, and repeating the acquisition with different filters (F1140C and F1550C). Cautioning about the very low jump detection threshold, they raised it from 4 to 8 in the F1140C filter. The detector reset anomaly was largely evaded in this study by removing the first integration of each exposure. "glow stick" stray light feature in the MIRI coronographic field of view was subtracted for each filter with a median background image of every 4-5 integrations from the dedicated background observations. They tried to extract the glow stick in a separate principal component with Karhunen-Loeve Image Processing (KLIP) and a Locally Optimized Combination of Images (LOCI), but either the residual flux was also degraded or variations between the integrations were not captured. Nonetheless, their median frame background subtraction worked very effectively, with not much room for improvement. For the absolute star position, several attempts were unsuccessful in pointing it, and the coronographic mask center was the proxy for the location. For both NIRCam and MIRI data, PSF was subtracted with three different PCA-based methods as ADI, RDI, and ADI+RDI. The companion HIP 65426 b was not detected only in the MIRI F1550C filter with ADI, whereas all other filters with ADI and RDI successfully detected it.

\centering\section{JWST Data Processing - Outline for Imaging}
\justifying

The following table and information are compiled from the webinar materials section of the Space Telescope Science Institute (STScI) website dedicated to \textit{JWebbinars} (STScI, 2022), and in the JWST User Documentation's \textit{calwebb\_detector1} (JWST User Documentation, 2022).
\begin{center}
\begin{tabular}{c c c c}
\hline
Stage & Step & NIRCam & MIRI\\
Stage 1 & Data Quality Initialization & \checkmark & \checkmark\\
Stage 1 & Saturation Check & \checkmark & \checkmark \\
Stage 1 & Error Initialization (EI) & \checkmark & \checkmark \\
Stage 1 & EI - Reset Anomaly Correction & & \checkmark \\
Stage 1 & EI - First-Frame Correction & & \checkmark \\
Stage 1 & EI - Last-Frame Correction & & \checkmark \\
Stage 1 & EI - Linearity Correction & & \checkmark \\
Stage 1 & EI - RSCD Correction & & \checkmark \\
Stage 1 & EI - Dark Subtraction & & \checkmark \\
Stage 1 & EI - Superbias Subtraction & \checkmark & \\
Stage 1 & Reference Pixel Correction (RPC) & \checkmark & \checkmark \\
Stage 1 & RPC - Linearity Correction & \checkmark & \\
Stage 1 & RPC - Persistence Correction & \checkmark & \\
Stage 1 & RPC - Dark Subtraction & \checkmark & \\
Stage 1 & Jump Detection & \checkmark & \checkmark \\
Stage 1 & Slope Fitting & \checkmark & \checkmark \\
Stage 2 & WCS Information & \checkmark & \checkmark\\
Stage 2 & Background Subtraction & \checkmark & \checkmark \\
Stage 2 & Flat Field Correction & \checkmark & \checkmark \\
Stage 2 & Flux Calibration & \checkmark & \checkmark \\
Stage 2 & Rectified 2D Product & \checkmark & \checkmark \\
Stage 3 & Refine WCS & \checkmark & \checkmark \\
Stage 3 & Moving Target WCS & \checkmark & \checkmark \\
Stage 3 & Background Matching & \checkmark &  \checkmark \\
Stage 3 & Outlier Detection & \checkmark & \checkmark \\
Stage 3 & Imaging Combination & \checkmark & \checkmark \\
Stage 3 & Source Catalog & \checkmark & \checkmark \\
Stage 3 & Updated Exposure Level Products & \checkmark &  \checkmark \\
\hline
\end{tabular}
\end{center}

Here is an overview of what these steps are and where can the above-mentioned data transformation methods enter the scene:\\
\textit{Stage 1}:\\
Data Quality Initialization: Pixels will have data quality flags and uncertainties which will be updated when necessary through the pipelines. \\
Saturation: This step aims to find pixels that encountered too many photons from extremely bright sources so after a threshold the incoming photons do not change the \textit{digital number} of the detector. Before finding the detector response for these photons, this step removes the saturated pixels.\\
Reset Anomaly Correction: This MIRI-specific correction is related to the anomaly in the first 12 groups in MIR ramps because of resetting in the dark. Reset anomaly is a systematic, additive, and transient offset.\\
First-frame Correction: MIRI's first frame has yet an uncharacterized transient feature. Hence, currently, it is being removed from the analysis.\\
Last-frame Correction: Similar to the first frame, the last frame, too, has an anomaly currently discarded from the further steps of the process.\\
Linearity Correction (MIRI): A well-characterized non-linear response of MIT detectors has a low-order polynomial structure. It is wavelength-dependent around >21 $\mu$m. Calibration reference files specific to the detector and wavelength account for this error.\\
Reset-switch charge decay - RSCD: A well-characterized decaying exponential proportional to the counts in the previous integration describes this integration-dependent transient. Thanks to detector-specific calibration reference files, this correction subtracts this exponential from the second and higher integrations. \\
Dark subtraction (MIRI): MIRI response has an integration number-dependent component, and a group-by-group dark subtraction removes this with integration-specific calibration reference files. \\
Superbias Subtraction: A specific bias offsets the digital number of each pixel and group, and this bias is proportional to the nonlinearity in a ramp. Calibration reference files assist in the removal of this error.\\
Linearity Correction (NIRCam): After correcting the superbias, the nonlinearity in the response of NIR detectors is captured by a low-order polynomial fit with detector-specific calibration reference files. \\
Persistence Correction: There is a persistent after-image appearance of previous exposures in the consecutive exposure in NIR detectors. It also has an exponentially-decaying change from one exposure to another. This error is corrected with detector-specific calibration reference files. \\
Dark Subtraction (NIRam): Excess response in the NIR detections in dark exposures are subtracted group-by-group with detector-specific calibration reference files. \\
Reference Pixel Correction (RPC): This is the calibration of pixels by an equivalent pixel insensitive to light so that electronic-related contributions to the response can be apparent and subtracted.\\
Jump Detection: When two consecutive detector groups have a larger difference in response than other consecutive detectors, the reason for this difference is generally cosmic rays hitting the corresponding pixels. The threshold to consider a value \textit{jump} is selected in terms of the number of standard deviations above the noise. \textit{Snowballs and showers} noise detection is also in the scope of this step.\\
Slope fitting: Also known as ramp fitting, it is essentially finding the response/photon slope after considering jump-flagged pixels and previous steps in the pipeline. \\
\textit{Stage 2}:\\
World Coordinate System (WCS) information: Instrument and detector-specific calibration reference files provide this WCS information and distortion model to transfer the pixel coordinates to astronomical right-ascension (RA) and declination (DEC) coordinates.\\
Background Subtraction: When a specific background target is available for the science target exposure, its combined background image is subtracted from the target exposure. \\
Flat-field Correction: This step corrects the pixel-to-pixel and large-scale variations in the response by the instrument- and detector-specific calibration reference files.\\
Flux Calibration: Flux calibration (or Photometric Calibration) apply a conversion factor between counts per second and Mega Jansky per steradian to the data; then, another conversion factor between counts/s and micro-Jansky/square arcsec is also present. \\
Rectified 2D product: This creates a rectified 2D image for visual inspection. \\
\textit{Stage 3}:\\
Refine Relative WCS - Tweakreg: \textbf{AstroDrizzle} utilizes common sources' locations between different images to refine their relative WCS information. \\
Moving target WCS: For moving targets, the final WCS of the output frame will be centered at the average location of the moving target with this step.\\
Background Matching - Skymatch: This step matches the background responses in different images by \textbf{AstroDrizzle} when the background overlaps. \\
Outlier Detection: This is a second check for outliers, now with \textbf{AstroDrizzle} code.\\
Imaging Combination - Resampling: This step mosaics multiple images in a specific band in an association file, exclıuding "bad" pixels. \\
Source Catalog: This step catalogs different sources of light in a given image with Photutils source extraction. There is also a deblending procedure to separate overlapping sources. \\
Update Exposure Level Products: It generates the highest-quality products after the pipeline.\\

\centering\subsection{Steps to Further Examine with Data Transformation Methods}
\justifying

Current approaches to reducing the raw JWST data continuously get better. Several of the steps in these pipelines might perform even more successfully with PCA-based data transformation methods.\\
\textit{Persistence Correction (Stage-1 Reference Pixel Correction)}: The image contain the actual exposure and respose(s) from previous exposure(s). An approach to disentangle these exposures from each other depending on the order of exposure might be tried with PCA and ICA, and \textit{a priori} number of components would simply be the order of the exposure.\\
\textit{Background Subtraction (Stage-2)}: The impact of the background subtraction step over the result of PCA; i.e. the difference in the results of principal components before and after the background subtraction step may uncover possible over-subtraction.\\
\textit{Background Matching - Skymatch (Stage-3)}: Before this step, the algorithm has single or multiple images containing signals from both the target and other point and diffuse sources, and sky. This is essentially a source-separation problem. There should be \textit{always present} uncorrelated and statistically-independent signal sources in every image, besides errors coming from cosmic rays, etc. Different exposures for a frame of a specific region might be binned as if different dimensions and all pixel values for the same pixel should be tried to be reduced to a lower number of \textit{components}. This may aid in the subtraction of the background.\\
\textit{Source Catalog (Stage 3)}: This part is normally with the Photutils package of Astropy in Python by using a threshold to find \textit{pixel values from a single source}. A pixel-based PCA-like method accounting for temporal variations between different integrations (such as SSA), might uncover more source-specific signals, or \textit{unmixing} contrarily or in addition to the segmentation-based approach. 

\centering\section{Concluding Remark}
\justifying

To sum up, there are many different steps that the advanced data transformation methods might yield more data-driven insight. Such potential rooms for improvements need to be probed without neglecting the computational efficiency and robustness of the potentially enhanced pipelines. 

\centering\section{References}
\justifying

\hangindent=0.7cm 
A. Claret. A new non-linear limb-darkening law for LTE stellar atmosphere models. Calculations for -5.0 <= log[M/H] <= +1, 2000 K <= Teff <= 50000 K at several surface gravities. \textit{Astronomy \& Astrophysics} \textit{363}. (2000).

\noindent \hangindent=0.7cm A. L. Carter, S. Hinkley, J. Kammerer, A. Skemer, B. A. Biller, J. M. Leisenring, M. A. Millar-Blanchaer, S. Petrus, J. M. Stone, K. Ward-Duong, J. J. Wang, J. H. Girard et al. The JWST Early Release Science Program for Direct Observations of Exoplanetary Systems I: High Contrast Imaging of the Exoplanet HIP 65426 b from 2-16 $\mu$m. (2022). arXiv: 2208.14990

\noindent \hangindent=0.7cm B. Everitt, T. Hothorn. An Introduction to Applied Multivariate Analysis with R. \textit{Springer}. (2011).

\noindent \hangindent=0.7cm E. M. Ahrer, K. B. Stevenson, M. Mansfield, S. E. Moran, J. Brande, G. Morello, C. A. Murray, N. K. Nikolov, D. J. M. Petit dit de la Roche, . Schlawin, P. J. Wheatley, et al. Early Release Science of the exoplanet WASP-39b with JWST NIRCam. (2022). arXiv: 2211.10489

\noindent \hangindent=0.7cm E. Schlawin, T. Beatty, B. Brooks, N. K. Nikolov, T. P. Greene, N. Espinoza, K. Glidic, K. Baka, E. Egami, J. Stansberry, M. Boyer et al. JWST NIRCam Defocused Imaging: Photometric Stability Performance and How it Can Sense Mirror Tilts. (2022). arXiv: 2211.16727

\noindent \hangindent=0.7cm G. Morello, A. Claret, M. Nartin-Lagarde, C. Cossou, A. Tsiaras, P.-O. Lagage. The ExoTETHyS Package: Tools for Exoplanetary Transits around Host Stars. \textit{The Astronomical Journal} \textit{159} (2020). doi: 10.3847/1538-3881/ab63dc

\noindent \hangindent=0.7cm H. F. Kaiser. The varimax criterion for analytic fotation in factor analysis. \textit{Psychometrika} \textit{23}. (1958). doi: 10.1007/BF02289233

\noindent \hangindent=0.7cm H. Hassani. A Brief Introduction to Singular Spectrum Analysis. Retrieved from: \url{https://ssa.cf.ac.uk/ssa2010/a_brief_introduction_to_ssa.pdf} (2010).

\noindent \hangindent=0.7cm H. Scharr. Optimale Operatoren in der Digitalen Bildverarbeitung. \textit{PhD Thesis}. (2000). doi: 10.11588/heidok.00000962

\noindent \hangindent=0.7cm H. Zou, T. Hastie, R. Tibshirani. Sparse Principal Component Analysis. \textit{Journal of Computational and Graphical Statistics} \textit{15}. (2006). doi: 10.1198/106186006x113430

\noindent \hangindent=0.7cm I. T. Joliffe. Principal Component Analysis. \textit{Springer}. 2nd Edition. (2002).

\noindent \hangindent=0.7cm J. A. Patel, N. Espinoza. Empirical Limb-darkening Coefficients and Transit Parameters of Known Exoplanets from TESS. \textit{The Astronomical Journal} \textit{163}. (2022). doi: 10.3847/1538-3881/ac5f55

\noindent \hangindent=0.7cm J. Bouwman, S. Kendrew, T. P. Greene, T. J. Bell, P. Lagage, J. Schreiber, D. Dicken, G. C. Sloan, N. Espinoza, S. Scheithauer, A. Coulais, O. D. Fox, et al. Spectroscopic time series performance of the Mid-Infrared Instrument on the JWST. (2022). arXiv: 2211.16123

\noindent \hangindent=0.7cm J. Canny. A Computational Approach to Edge Detection. \textit{IEEE Transactions on Pattern Analysis and Machine Intelligence}. \textit{PAMI-8} \textbf{679}. (1986). doi: 10.1109/TPAMI.1986.4767851

\noindent \hangindent=0.7cm J. H. Girard, J. Leisenring, J. Kammerer, M. Gennaro, M. Rieke, J. Stansberry, A. Rest, E. Egami, B. Sunnquist, et al. JWST/NIRCam Coronography: Commissioning and First On-Sky Results. (2022). arXiv: 2208.00998

\noindent \hangindent=0.7cm J. Shlens. A Tutorial on Independent Component Analysis. (2014b). arXiv: 1404.2986

\noindent \hangindent=0.7cm J. Shlens. A Tutorial on Principal Component Analysis. (2014). arXiv:1404.1100

\noindent \hangindent=0.7cm JWST User Documentation. calwebb\_detector1. \url{https://jwst-docs.stsci.edu/jwst-science-calibration-pipeline-overview/stages-of-jwst-data-processing/calwebb\_detector1}. (2022).

\noindent \hangindent=0.7cm L. Carone, P. Mollière, Y. Zhou, J. Bouwman, F. Yan, R. Baeyens, D. Apai, N. Espinoza, B. V. Rackham, A. Jordán. Indications for very high metallicity and absence of methane in the eccentric exo-Saturn WASP-117b. \textit{Astronomy \& Astrophysics} \textit{646}. (2021). doi: 10.1051/0004-6361/202038620

\noindent \hangindent=0.7cm L. Kreiderg. \textbf{batman}: BAsic Transit Model cAlculatioN in Python. \textit{Publications of the Astronomical Society of the Pacific} \textit{127}. (2015). doi: 10.1086/683602

\noindent \hangindent=0.7cm N. Astudillo-Defru, R. Cloutier, S. X. Wang, J. Teske, R. Brahm, C. Hellier, G. Ricker, R. Vanderspek, D. Latham, S. Seager, J. N. Winn, J. M. Jenkins, K. A. Collins, K. G. Stassun, et al. A hot terrestrial planet orbiting the bright M dwarf L 168-9 unveiled by TESS. \textit{Astronomy \& Astrophysics} \textit{636}. (2020). doi: 10.1051/0004-6361/201937179

\noindent \hangindent=0.7cm N. Golyandina, A. Korobeynikov. Basic Singular Spectrum Analysis and Forecasting with R. (2013). arXiv: 1206.6910

\noindent \hangindent=0.7cm N. Golyandina, V. Nekrutkin, A. Zhigljavsky. Analysis of Time Series Structure: SSA and Related Techniques. Chapmann\&Hall/CRC. (2001).

\noindent \hangindent=0.7cm N. Nadisic, A. Vandaele, J. E. Cohen, N. Gillis. Sparse Separable Nonnegative Matrix Factorization. (2006). arXiv: 2006.07553

\noindent \hangindent=0.7cm R. Li, H. Li, F. Wang. Dependent Component Analysis: Concepts and Main Algorithms. \textit{Journal of Computers} \textit{5}. (2010). doi: 10.4304/jcp.5.4.589-597

\noindent \hangindent=0.7cm STScI. Learning JWST Data Analysis With the JWebbinars. \textit{Materials \& Videos}, \textit{3-Pipeline: Imaging Mode (May 2021)}. \url{https://stsci.app.box.com/s/gz5q1r8ijb1hvimmeh8fi7xlb3b368ds}. (2022).

\noindent \hangindent=0.7cm T. J. Bell, E. M. Ahrer, J. Brande, et al. Eureka!: An End-to-End Pipeline for JWST Time-Series Observations. (2022). arXiv: 2207.03585

\end{document}